\documentclass[12pt,final]{elsarticle}

\usepackage{mathptmx}
\usepackage[numbers]{natbib}
\usepackage{hyperref}
\usepackage{amsmath}
\usepackage{amsfonts}
\usepackage{amsthm}
\usepackage{amssymb}
\usepackage{color}
\usepackage{graphicx}
\usepackage{multirow}

\numberwithin{equation}{section}

\begin{document}

\begin{frontmatter}


   \title{Response to Comments in ``Exact and ``exact" formulae in the theory of composites" (arXiv:1708.02137v1 [math-ph], 7 August 2017)}

  \author[UFRJ]{Manuel Ernani Cruz\corref{cor}}\cortext[cor]{Corresponding
    author.}  \ead{manuel@mecanica.coppe.ufrj.br}
	\author[Matcom]{Juli\'an Bravo-Castillero} \ead{jbravo@matcom.uh.cu} 	

  \address[UFRJ]{UFRJ-Federal University of Rio de Janeiro, PEM/COPPE, CP 68503, Rio de Janeiro, RJ, 21941-972, Brazil}
  \address[Matcom]{Facultad de Matem\'{a}tica y Computaci\'{o}n, Universidad de La Habana, San Lazaro y L, Vedado, CP 10400, Cuba}

\begin{abstract}
In this paper we present our response to the comments by Andrianov and Mityushev regarding a recent publication of ours on the determination of the effective thermal conductivity of multiscale ordered arrays.
\end{abstract}

\end{frontmatter}

In their recent publication, Andrianov and Mityushev \cite{AM} discuss the proper utilization of the terms analytical formula, approximate solution, closed form solution, asymptotic formula and others in the context of methods devoted to the determination of effective properties of composite materials.

Since the pioneering works by Lord Rayleigh \cite{R} and Maxwell \cite{M} it is verified that the determination of effective properties of heterogeneous materials \cite{T} challenges researchers active in diverse fields, spanning physics, mathematics, and engineering disciplines. As a consequence, a wide range of methods and techniques have been developed using different building blocks. It is, therefore, not unexpected that differences in notations and in the usage of some terms are bound to occur.

In their publication, Andrianov and Mityushev \cite{AM} makes several comments on our recent paper \cite{NCC} concerned with the calculation of the effective thermal conductivity of three-scale arrangements of circular cylinders and spheres orderly arranged, respectively, in the 2-D square and 3-D simple cubic arrays. For that matter, we would like to thank Andrianov and Mityushev for their interest in our paper.

In Ref. \cite{NCC}, the inclusions (circular cylinders or spheres) are periodically distributed throughout two microstructural levels of disparate length scales, such that the ratio of the radii of two inclusions in the small $z$-scale microstructural level and in the intermediate $y$-scale level is infinitesimally small for finite concentrations. To characterize the macroscopic behavior of such arrays through the calculation of their effective conductivities, the present authors combine the analytical results derived in \cite{RCC} from application of the reiterated homogenization method \cite{BP,BLP} to the multiscale heat conduction problem with known algebraic formulae \cite{MT,SA} for the effective conductivities of the respective individual two-scale (monodisperse) arrays.

The first comment by Andrianov and Mityushev regards the sentences ``The interaction of periodic multiscale heterogeneity arrangements is exactly accounted for by the reiterated homogenization method. The method relies on an asymptotic expansion solution of the first principles applied to all scales, leading to general rigorous expressions for the effective coefficients of periodic heterogeneous media with multiple spatial scales." used in Ref. \cite{NCC} to indicate one of the strengths of the adopted approach. Andrianov and Mityushev \cite{AM} affirm that ``This declaration is not true, because a large particle does not interact with another particle of vanishing size." It is remarked, that the present authors do not declare in Ref. \cite{NCC} that a large particle interacts with another particle of vanishing size. Certainly, there are thermal interactions of particles in the intermediate $y$-scale microstructural level, and there are interactions of particles in the small $z$-scale microstructural level. In addition, there is a thermal interaction of the whole small-scale microstructural level with the whole intermediate-scale microstructural level, such that the effective conductivity of the three-scale medium changes relative to that of the two-scale medium, as demonstrated in Ref. \cite{NCC}. The reiterated homogenization method does exactly account for all these interactions, in the asymptotic limits required by the theory \cite{BP,BLP}. In the light of this second interpretation, the original sentences in Ref. \cite{NCC} are, indeed, true.

The second comment by Andrianov and Mityushev regards the fact that, in their view, an effective medium approximation valid for dilute composites was actually applied in Ref. \cite{NCC} as the reiterated homogenization theory. Andrianov and Mityushev \cite{AM} then conclude that ``As a consequence, formulae (18) and (19) from \cite{NCC} can be valid to the second order of concentration and additional numerically calculated ``terms" are out of the considered precision" (to avoid confusion here and elsewhere, we use the reference numbers in the reference list of this present paper, rather than those used in Ref. \cite{AM}). In Ref. \cite{NCC}, the present authors do not apply effective medium approximations in lieu of reiterated homogenization theory. Instead, it is first observed in Ref. \cite{NCC} that the expression for the effective thermal conductivity of a two-scale periodic medium obtained by conventional homogenization \cite{BP,BLP} is similar to the expressions for the effective conductivities of a three-scale periodic medium obtained by reiterated homogenization \cite{RCC,BLP}. This similarity allowed the present authors to derive equation (16) in Ref. \cite{NCC}, which, together with Eqs. (8) and (10) of the same reference, permit the physically meaningful comparison of the effective conductivity of a two-scale (monodisperse) heterogeneous medium with that of a three-scale (bidisperse) heterogeneous medium with similar parameters. Specifically, to carry out the comparison with numerical values of effective conductivity, as pointed out in the beginning of Section 3 in \cite{NCC}, computations were performed using well-known algebraic formulae available in the literature \cite{MT,SA} for two-scale media, thus valid for each microstructural level of the considered three-scale media, in view of the mentioned similarity. Although the formulae (18) and (19), obtained from [9], and also formulae (20) and (21), obtained from [10], used in Ref. \cite{NCC} have limited precision, as Andrianov and Mityushev \cite{AM} point out, they are not restricted to the dilute regime of concentration for composites. In fact, terms up to the 7th order in concentration are included in formulae (18)-(19), and terms up to the 9th order in concentration are included in formulae (20)-(21). These attributes of the formulae were noted in Ref. \cite{NCC}, where the objective was to demonstrate the gains in effective conductivity obtained with (bidisperse) three-scale arrays relative to the (monodisperse) two-scale counterparts.

Lastly, the third comment by Andrianov and Mityushev \cite{AM} states that ``The considered problem refers to the general polydispersity problem discussed in 2D statement in \cite{BM}. It is not surprisingly that the description of the polydispersity effects in \cite{BM} and \cite{NCC} are different since the ``exact" formula from \cite{NCC} holds only in the dilute regime. This is the reason why the effective conductivity is less than the conductivity of matrix reinforced by higher conducting inclusions in Figs. 3 and 7 of \cite{NCC} for high concentrations." Again, as remarked before, the formulae (18)-(19) and (20)-(21) used in Ref. \cite{NCC} hold beyond the dilute regimes of the respective square and cubic arrays \cite{MT,SA}. Interestingly, it is possible to view the study in Ref. \cite{NCC} as dealing with a polydispersity problem. However, the problem considered in Ref. \cite{NCC} is a specific limiting case of polydispersity, namely, that due to periodic inclusions located in different microstructural levels of very disparate length scales. It is apparent that this particular polydispersity two-level problem does not fit into the description of the one-level media dealt with in Ref. \cite{BM}. In particular, the three-scale periodic arrays obey the volume fraction formula (8) in Ref. \cite{NCC}. It is not possible to apply formulae (5) and (6), and the microstructure depicted in Fig. 2 in Ref. \cite{BM} to geometrically describe the three-scale periodic arrays. Therefore, the study in Ref. \cite{BM} is not directly applicable to the study in Ref. \cite{NCC}.

Figs. 3 and 7 in Ref. \cite{NCC} show the gains in the effective conductivities of the (bidisperse) three-scale arrays relative to the corresponding (monodisperse) two-scale arrays, not the gains relative to the matrix. As also obtained in \cite{BM}, depending on the problem volume fractions, the bidisperse three-scale-array effective conductivity can increase or decrease relative to the monodisperse two-scale-array effective conductivity, and thus can display the S-shape behavior pointed out by Andrianov and Mityushev \cite{BM}. However, beyond a certain bulk volume fraction, the three-scale-array effective conductivity is less than the two-scale-array effective conductivity for the whole range of the small $z$-scale volume fraction. As pointed out in Ref. \cite{NCC}, this behavior occurs when the path for heat transfer no longer improves, as one removes material and reduce the size of the inclusion at the intermediate $y$-scale, and distribute the material as inclusions at the small $z$-scale. This is physically plausible for the ordered two-level arrays considered in Ref. \cite{NCC}, for which the relations (63) for the bumping model in \cite{BM} are not verified due to the different laws of formation for the microstructures.

\vspace{0.5cm}

{\bf Acknowledgments}

\vspace{0.3cm}

MEC thanks the support provided by CNPq-Brazilian National Council for Scientific and Technological Development. JB acknowledges the C\'atedra Extraordinaria IIMAS and  PREI-DGAPA, UNAM.

\end{document}